\documentclass[12pt]{article}

\usepackage{epsf}
\usepackage[dvips]{graphicx}

\begin{document}

\begin{center}
{\bfseries LONG-RANGE RAPIDITY CORRELATIONS \\
IN THE MODEL WITH INDEPENDENT EMITTERS}

\vskip 5mm

V.V. Vechernin

\vskip 5mm

{\small
{\it
St.-Petersburg State University
}
\\
{\it
E-mail: vechernin@pobox.spbu.ru
}}
\end{center}

\vskip 5mm

\begin{center}
\begin{minipage}{150mm}
\centerline{\bf Abstract}
The correlation between multiplicities
in two separated rapidity windows,
the so-called long-range correlation (LRC),
is studied
in the framework of the model with independent identical emitters.
It's shown that the LRC coefficient, defined for the scaled (relative)
variables, nevertheless depends on the absolute width of the forward rapidity window
and does not depend on the width of the backward one.
The dependence of the LRC coefficient on the forward rapidity acceptance
is explicitly found with only one theoretical parameter.
The preliminary comparison with ALICE 7~TeV pp collisions data
shows that the multiplicity LRC in the data
can be described in the framework of the suggested approach.
\end{minipage}
\end{center}

\vskip 10mm

\def\bc{\begin{center}}
\def\ec{\end{center}}
\def\beq{\begin{equation}}
\def\eeq{\end{equation}}
\def\noi{\noindent}
\def\hs#1{\hspace*{#1cm}}
\def\av#1{\langle #1 \rangle}
\def\avr#1#2{\langle {#1} \rangle^{}_{#2}}
\def\DF{{\Delta y^{}_F}}
\def\DB{{\Delta y^{}_B}}
\def\DphiF{{\Delta \varphi^{}_F}}
\def\DphiB{{\Delta \varphi^{}_B}}

\def\nF{{n_F^{}}}
\def\pF{{p_{tF}^{}}}
\def\nB{{n_B^{}}}
\def\pB{{p_{tB}^{}}}
\def\mF{{\mu^{}_F}}
\def\mB{{\mu^{}_B}}

\def\onF{{\overline{n}_F^{}}}
\def\onB{{\overline{n}_B^{}}}
\def\omF{{\overline{\mu}^{}_F}}
\def\omB{{\overline{\mu}^{}_B}}

\def\moF{{\mu^{}_{0F}}}
\def\moB{{\mu^{}_{0B}}}

\def\onnF{{\overline{n_F^2}}}
\def\onnB{{\overline{n_B^2}}}
\def\onFF{{\overline{n}_F^2}}
\def\onBB{{\overline{n}_B^2}}

\def\ommF{{\overline{\mu_F^2}}}
\def\ommB{{\overline{\mu_B^2}}}
\def\omFF{{\overline{\mu}_F^2}}
\def\omBB{{\overline{\mu}_B^2}}
\def\oF{{\overline{F}}}
\def\oB{{\overline{B}}}
\def\ooF{{\overline{F_{}^2}}}
\def\ooB{{\overline{B_{}^2}}}
\def\oFF{{\overline{F}_{}^2}}
\def\oBB{{\overline{B}_{}^2}}

\def\nFi{n_{F}^i}
\def\nBi{n_{B}^i}
\def\nFj{n_{F}^j}
\def\nBj{n_{B}^j}
\def\df{\delta_{F,\sum F_i}}
\def\db{\delta_{B,\sum B_i}}
\def\pp{\prod_{i=1}^N p(B_i,F_i)}
\def\sumn{\sum^n_{i=1}}

\def\oN{\overline{N}}
\def\obr{\overline{b}^{rel}_{}}
\def\br{b^{rel}_{}}
\def\ba{b^{abs}_{}}
\def\oba{\overline{b}^{abs}_{}}
\def\ar{a^{rel}_{}}
\def\aa{a^{abs}_{}}
\def\dnF{\nF-\av{\nF}}

\section{Introduction}

In processes of the multiple production
in pp and AA collisions at high energies
one can study the correlation between multiplicities $n_F$ and $n_B$
of charged particles in two rapidity windows (``forward'' and ``backward'')
separated by some gap -
the so-called long-range rapidity correlation (LRC).
In present paper we consider this correlation
in the framework of the model with independent identical emitters.

To analyze the correlation one usually introduces
the correlation function (regression) $f(\nF)\equiv\av{\nB}_{\nF}^{}$
and studies the mean multiplicity in the backward window
as a function of the multiplicity in the forward window.
In the case of linear regression
one defines the correlation coefficient $\ba$,
characterizing a strength of the correlation,
by the following way:
\begin{equation}
\label{f_abs_lin}
\av{\nB}_{\nF}^{}= \aa+\ba \nF \ .
\end{equation}
But the value of such defined correlation coefficient
obviously depends on the lengths of forward $\DF$ and
backward $\DB$ rapidity windows,
because the $n_F$ and $n_B$ depends on these lengths.

To eliminate this trivial dependence on the widths of rapidity windows
we define the correlation
coefficient $\br$ using the scaled (relative) variables:
\begin{equation}
\label{f_rel_lin}
\frac{\av{\nB}_{\nF}^{}}{\av{\nB}}
= \ar+\br\  \frac{\nF-\av{\nF}}{\av{\nF}}
= \ar+\br \left(\frac{\nF}{\av{\nF}}-1\right) \ .
\end{equation}
Clear that $\ba$ and $\br$ are simply connected $\br = \frac{\av{\nF}}{\av{\nB}}\ba$
and for symmetric $\DB=\DF$ windows   $\av\nB=\av\nF$ and  $\br =  \ba $.

For nonlinear correlation function $f(\nF)$
it seems reasonable \cite{Vestn1}-\cite{PPR}
to define the correlation coefficients as follows
\begin{equation}
\label{def1}
\textsf{Def.1:}\hs1 \ba \equiv \left.\frac {d\av{\nB}_{\nF}^{}} {d\nF}
\right|_{\nF=\av{\nF}} \ , \hs1
\br \equiv \left.\frac {d\av{\nB}_{\nF}^{}/\av{\nB}} {d\nF/\av{\nF}}
\right|_{\nF=\av{\nF}}
= \frac{\av{\nF}}{\av{\nB}}\ \ba \ .
\end{equation}

\section{Model}

In the model with independent identical emitters \cite{PLB00} one assumes
that the probability $P(\nB,\nF)$ to observe simultaneously
the $n_F$ charged particles in the forward rapidity window
and  the $n_B$ particles in the backward one
is given by the expression:
\begin{equation}
\label{P_BF}
P(B,F)=\sum_N w(N)\sum_{B_1,...,B_N}\sum_{F_1,...,F_N}{\delta_{B\ B_1+...+B_N}}
{\delta_{F\ F_1+...+F_N}}
\pp  \ ,
\end{equation}
where we have used short notations:
$$
F\equiv\nF \ , \hs1  B\equiv\nB\ ; \hs1  F_i\equiv\nFi\ , \hs1  B_i\equiv\nBi
$$
In the formula (\ref{P_BF}) the $w(N)$ is the probability to have $N$ emitters
in the given event and the $p(B_i,F_i)$ is the probability
that $i$-th emitter produces
the $F_i$ charged particles in the forward rapidity window
and the $B_i$ particles in the backward one.

In the case of long-range correlations (LRC)
with a sufficiently large rapidity gap between windows
the model supposes that every emitter (string) produces particles  \emph{independently}
in the forward and backward windows:
\beq \label{factor}
p(B_i,F_i)=p_B(B_i)\ p_F(F_i) \  .
\eeq
So the correlation arises only due to event-by-event fluctuations
of the number of emitters.

In paper \cite{Vestn1}
using methods developed in \cite{PLB00}
for various distributions in some approximation
the following formula
for the defined (\ref{def1}) correlation coefficient $\br$
was obtained:
\begin{equation}
\label{br}
\br=\frac{\kappa\, \omF}{\kappa\, \omF+1} \ .
\end{equation}
Here the $\kappa$ is the ratio of two scaled variances:
\begin{equation}
\label{kappa}
\kappa=\frac{V_N}{V_\mF}\ , \hs 1
V_N=\frac{D_N}{\av N}\ , \hs 1
V_\mF=\frac{D_\mF}{\omF}\ ,
\end{equation}
where $\av N$ and $D_N= \av {N^2}-{\av N}^2$ are the mean number of emitters
and the event-by-event variance of the number of emitters.
The $\omF$ and $D_\mF=\overline{\mu^2_F}-\overline{\mu}^{2}_F$ are the mean multiplicity
produced by \emph{one} emitter in the \emph{forward} window and the corresponding variance.
For Poisson distributions, for example,  $V_N=V_\mF=1$ and $\kappa=1$,
then the $\br$ (\ref{br}) depends only on $\omF$.

Clear that the $\omF$ depends
on the width of the forward rapidity window.
For the forward window in the plateau region one can assume
\begin{equation} \label{mu0}
\omF=\moF \DF
\end{equation}
where  $\moF$  is the average multiplicity produced by \emph{one} emitter in
the \emph{forward} window
per a \emph{unit} of rapidity. Then by (\ref{br}) we have for the correlation coefficient:
\beq \label{br_dyF}
\br=\frac{\kappa\mu_{0F}\, \DF}{\kappa\mu_{0F}\, \DF+1}
=\frac{a\, \DF}{a\, \DF+1} \ ,
\eeq
where $a=\kappa\mu_{0F}$ is the only theory parameter.

Note that in the case of limited
azimuth acceptance $\DphiF$ in the forward rapidity window
one has to use the formulas
\begin{equation} \label{DphiF}
\omF=\moF \DF \DphiF/2\pi \ , \hs1
\br=\frac{a\, \DF\DphiF/2\pi}{a\, \DF\DphiF/2\pi+1}
\end{equation}
instead of (\ref{mu0}) and (\ref{br_dyF}).

So we see from formula (\ref{br}) that the multiplicity LRC coefficient $\br$
even defined  for \emph{scaled
variables} (\ref{def1}) nevertheless depends through $\mF$
on the length and the azimuth acceptance
of the \emph{forward} rapidity window $\Delta y_F$, $\DphiF$ and
does not depend on the length and the azimuth acceptance
of the backward one $\Delta y_B$, $\DphiB$.
The reason is that the regression procedure is being made by the forward window.
One can find the physical discussion of this phenomenon in ref. \cite{Vestn1}.

\section{Alternative definition}

In some papers instead of the definition (\ref{def1})
the following definition of the correlation coefficient is used
\beq \label{def2}
\textsf{Def.2:} \hs1
\ba = \frac{\av{\nB\nF}-\av{\nB}\av{\nF}} {\av{n_F^2}-\av{\nF}^2_{}}
=\frac{\av{\nB\nF}-\av{\nB}\av{\nF}} {D_\nF}  \ , \hs1
\br = \frac{\av{\nF}}{\av{\nB}}\ \ba \ .
\eeq
For a linear correlation function $f(x)$
these formulae can be obtained by (\ref{f_abs_lin}) and (\ref{f_rel_lin}) exactly,
but in the case of a nonlinear correlation function
the definitions (\ref{def1}) and (\ref{def2})
are not identical (see discussion below).
Note that when one extracts the correlation coefficient from the experimental data
the definition 1 reduces to the definition 2 but with a \emph{narrow interpolation interval}
centered around $\nF=\av\nF$, instead of the whole $\nF$ range in (\ref{def2}).

Using the definition (\ref{def2}) one can obtain
the formula (\ref{br}) for the correlation coefficient
at very general assumptions,
because in this case instead of a calculation of
the correlation function $f(x)$
one needs to calculate
only some averages $\av{\nB\nF}$, $\av{n_F^2}$, $\av{\nF}$, $\av{\nB}$,
which is much more simple.

As an example let us to calculate $\av{n_F^2}$.
By (\ref{P_BF}) and (\ref{factor}) we have
\beq \label{nF2}
\av{n_F^2}\equiv\av{F^2_{}}\equiv\sum_{B,F} F^2 P(B,F)
=\sum_{F} F^2\  \sum_Nw(N)\sum_{F_1,...,F_N}
{\delta_{F\ F_1+...+F_N}}
\prod_{i=1}^N p_F(F_i)=
\eeq
$$
=\sum_Nw(N)\!\!\sum_{F_1,...,F_N}
(F_1+...+F_N)^2
\prod_{i=1}^N p_F(F_i)
=\sum_Nw(N)\!\!\sum_{F_1,...,F_N}
[\sum_{i=1}^N F^2_i+\!\sum_{i\neq j=1}^N F^{}_i F^{}_j]
\prod_{i=1}^N p_F(F_i)=
$$
$$
=\sum_N w(N) [N \ommF + (N^2-N) \omFF] = \av N \ommF + (\av {N^2}-\av N) \omFF
= \av N (\ommF-\omFF)+\av {N^2} \omFF \ ,
$$
where we have used that for identical emitters for any $i$:
$$
\sum_{F_i}F_i\ p_F(F_i)=\omF
 \ , \hs1
\sum_{F_i}F^2_i\ p_F(F_i)=\ommF  \ .
$$
So we obtain the well known formula for the variance $D_\nF$ in the denominator
of the (\ref{def2}):
\beq \label{DnF}
D_\nF\equiv\av{n_F^2}-\av{n_F^{}}^2 = \av N D_\mF +\av {N^2} \omFF - \av N^2 \omF^2=
\av N D_\mF +D_N \omFF
\eeq
Similarly one finds for the correlator $\av{\nB\nF}-\av{\nB}\av{\nF}$
in the numerator of the (\ref{def2}):
\beq \label{cor}
\av{\nB\nF}-\av{\nB}\av{\nF}=D_N \omB \omF
\eeq
Substituting (\ref{DnF}) and (\ref{cor}) in (\ref{def2})
we comes again to the expression (\ref{br})
for the correlation coefficient.

\section{Comparison of the definitions}

In the case of a nonlinear regression one can expand
the correlation function in powers of the deviation of the $\nF$ from its mean value $\av\nF$:
\begin{equation}
\label{ser}
\av{\nB}_{\nF}^{}\equiv f(\nF)=\sum_{k=0}^{\infty} (\dnF)^k f_k \ .
\end{equation}
Clear that by the first definition (\ref{def1}):
\begin{equation} \label{f1}
\ba=f_1 \ , \hs1 \br =  (\av{\nF}/\av{\nB})f_1  \ .
\end{equation}
If we now apply the second definition (\ref{def2}),
we get another expression
for the correlation coefficient:
\begin{equation}
\label{difr}
\ba=f_1+D^{-1}_\nF\sum_{k=2}^{\infty} \av{(\dnF)^{k+1}} f_k \ .
\end{equation}
Comparing (\ref{difr}) with (\ref{f1})
we see that the difference between these two definitions
depends on the higher moments of the $\nF$ distribution.

Consider also a constant $f_0$ in the expansion (\ref{ser}),
$f_0=f(\av\nF)=\av{\nB}_{\nF=\av\nF}^{}$.
In the case of a nonlinear correlation function
one can introduce the coefficient $\ar$ by the following way:
\beq \label{ar}
\ar=\frac{f_0}{\av{\nB}}=\frac{\av{\nB}_{\nF=\av\nF}^{}}{\av{\nB}}
=\frac{f(\av\nF)}{\av{f(\nF)}} \ .
\eeq
For a linear correlation function this definition coincides with (\ref{f_rel_lin}).
After some trivial manipulations using (\ref{ser}) one can get:
\begin{equation}
\label{ar_ser}
\ar=1 - \av\nB^{-1}_{} \left(D_\nF f_2 + \sum_{k=3}^{\infty} \av{(\dnF)^{k}} f_k\right) \ .
\end{equation}
It follows from this formula that
for linear correlation function: $\ar=1$.
In the next (quadratic) approximation we have: if $\ar>1$,
then the correlation function is convex upwards: $f_2<0$ and vice versa.

\section{Conclusion}

It's shown that the formula obtained in \cite{Vestn1} for the long-range multiplicity
correlation coefficient in the model with independent emitters:
$$
\br=\frac{\kappa\, \omF}{\kappa\, \omF+1} \ ,
$$
where the $\kappa$ is the ratio of two scaled variances:
$\kappa=V_N/V_\mF$, $V_N=D_N/\av N$,  $V_\mF=D_\mF/\omF$ and $\omF$ is the mean multiplicity
produced by \emph{one} emitter in the \emph{forward} window,
is valid at very general assumptions.

As a result the multiplicity correlation coefficient $\br$ defined for \emph{scaled
variables} nevertheless depends on the width of the \emph{forward} rapidity window
$\Delta y_F$ and does not depend on the width of the backward one $\Delta y_B$.
For example, for the forward window in the plateau region, when one can assumes
$\omF=\moF \DF$, we have
$$ \br=\frac{\kappa\mu_{0F}\, \DF}{\kappa\mu_{0F}\, \DF+1} \ ,
$$
where  $\mu_{0F}$  is the average multiplicity produced by \emph{one} emitter in
the \emph{forward} window
per a \emph{unit} of rapidity.
The same is valid for the forward azimuth acceptance - $\DphiF/2\pi$
(see formula (\ref{DphiF})).
The reason is that the \emph{regression procedure} is being made by the \emph{forward} window.
One can find the physical discussion of this phenomenon in ref. \cite{Vestn1}.

The preliminary comparison with ALICE 7 TeV pp collisions data
shows that the multiplicity LRC can be described in the framework of the suggested approach
at the value of the only theory parameter $a=\kappa\mu_{0F}=1.8$.
Note that the transverse momentum LRC are absent in the model
with independent identical emitters.
To describe them one has to take into account
the interaction of emitters (a string fusion or other collectivity effects).

The author thanks M.A.~Braun and G.A.~Feofilov for useful discussions.
The work was supported
by the RFFI grants 09-02-01327-a and 08-02-91004-CERN-a.

\end{document}